# Exploiting Environmental Computation in a Multi-Agent Model of Slime Mould.

Jeff Dale Jones

Centre for Unconventional Computing,
University of the West of England, Coldharbour Lane,
Bristol, BS16 1QY, UK.
jeff.jones@uwe.ac.uk

**Abstract.** *Very simple organisms, such as the single-celled amoeboid slime mould Physarum polycephalum possess no neural tissue yet, despite this, are known to exhibit complex biological and computational behaviour. Given such limited resources, can environmental stimuli play a role in generating the complexity of slime mould's behaviour? We use a multi-agent collective model of slime mould to explore a two-way mechanism where the collective's behaviour is influenced by simulated chemical concentration gradient fields and, in turn, this behaviour alters the spatial pattern of the concentration gradients. This simple mechanism yields complex behaviour amid the dynamically changing gradient profiles and suggests how the apparently `intelligent' response of the slime mould could possibly be due to outsourcing of computation to the environment.*

The true slime mould *Physarum polycephalum* is a single-celled organism possessing no nervous system yet (in its vegetative plasmodium stage) exhibits a complex range of biological and computational behaviours (for an overview of its abilities see [1]). Plasmodium of *P. polycephalum* is comprised of an adaptive gel/sol transport network. The ectoplasmic gel phase is composed of a sponge-like matrix of actin and myosin fibres through which the endoplasmic sol flows, transported by spontaneous and self-organised oscillatory contractions. The organism behaves as a distributed computing material, capable of responding to a wide range of spatially represented stimuli, including chemoattractants, chemorepellents, temperature changes and light irradiation. Slime mould performs these complex feats using only very simple components and its behaviour has even been described as intelligent [2]. How can such complex behaviour emerge in such a simple organism? In this abstract we describe a two-way mechanism by which environmental information in the form of spatial concentration gradients can be exploited to generate apparently intelligent spatial behaviours seen in slime mould.

We use a multi-agent particle based model of slime mould which replicates the self-organised network formation and adaptation of slime mould (see [3] for a detailed description). Coupled mobile agents move on a 2D lattice, sensing and depositing a simulated chemoattractant which diffuses over time. The virtual slime mould is comprised of a large population of these simple agents which collectively exhibits cohesion, network formation, network minimisation and shape adaptation. Simple rules (based on local space availability and crowding) govern the growth and shrinkage of the collective. Environmental stimuli in the form of attractants (positive displacement of the lattice) and repellents (negative displacement) are projected onto the same lattice and the diffusing stimuli cause the collective to deform at its border and move towards attractants and away from repellents.

**Concentration Dependent Morphology**

As observed in *P. polycephalum* [4], attractant concentration affects the growth patterns of the collective – low-concentration stimuli results in pseudopod-like growth fronts and dendritic networks (Fig. 1), whereas high-concentration stimuli generates florid radial growth patterns (Fig. 2).

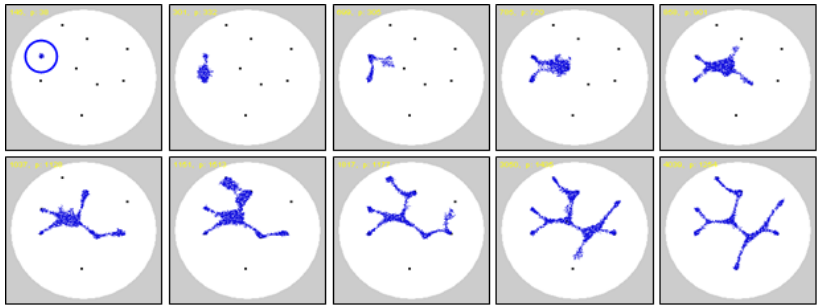

**Fig. 1:** Dendritic growth of virtual plasmodium in response to low-concentration stimuli. Inoculation site is circled, attractant stimuli indicated by dots, emergent transport network connects the attractant sources.

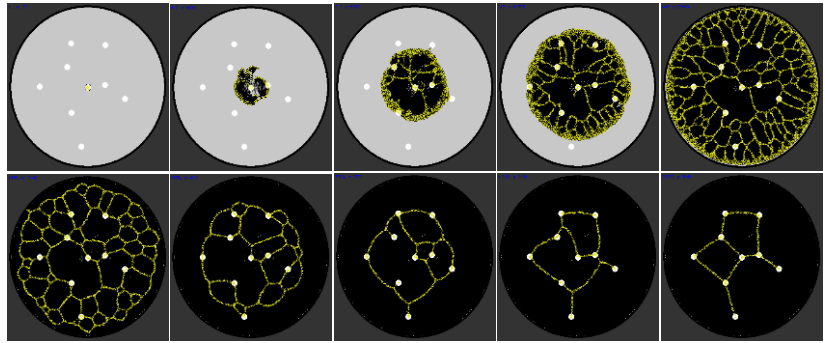

**Fig. 2:** Radial expansive growth of virtual plasmodium in response to high-concentration stimuli. Inoculation site is at the centre of the lattice on simulated nutrient-rich substrate (light grey) with individual 'oat flakes' (white). Network growth is radial and network minimisation occurs when the background substrate is depleted.

The different morphologies emerge because weaker concentration gradients present a smaller, isolated and directed stimulus region to the collective, whereas higher concentrations (for example within a nutrient-rich oatmeal agar substrate) present an all-encompassing stimulus, encouraging growth in all directions. The spatial pattern of concentration gradients presents a 2D stimulus map influencing the growth pattern and direction of the collective. These results suggest that the apparent complexity (or 'intelligence') of slime mould's behaviour may be partially due to the configuration of its environment.

**Two-way Interaction with Gradient Fields**

The virtual plasmodium does not just simply follow the concentration profile of this 2D stimulus map but also modifies it, by suppressing the fields via engulfment and consumption of nutrients. The result is a dynamical interaction between the collective and the environment which may be summarised (using only attractants, for simplicity) in the following scheme:

1. A chemoattractant gradient field is generated by the diffusion of nutrients.
2. The collective is attracted towards, and moves towards, the higher concentration gradient profiles.
3. On encountering the source of the attractant, the collective suppresses and/or consumes the nutrient, reducing its chemoattractant concentration gradient.
4. The spatial pattern of concentration gradients is changed, altering the stimulus presented to the virtual plasmodium.
5. Resume part 2.

The effect of this two-way interaction between environment (presentation of gradient) and virtual plasmodium (modification of gradient) can be seen when the concentration gradient is visualised, as shown in in Fig. 3 which demonstrates the construction of a simple spanning tree network by the virtual plasmodium.

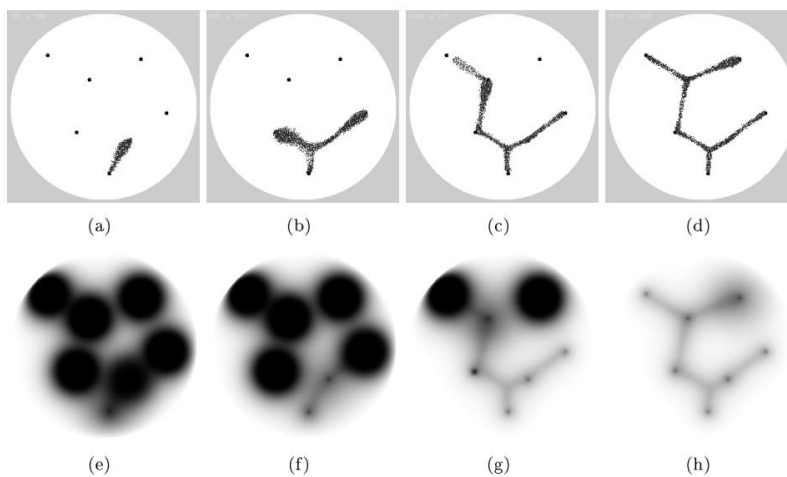

**Fig. 3:** Construction of a spanning tree by virtual plasmodium. (a) Small population (particle positions shown) inoculated on lowest node (bottom) growing towards First node and engulfing it, reducing chemoattractant projection, (b-d) Model population grows to nearest sources of chemoattractant completing construction of the spanning tree, (e-h) Visualisation of the changing chemoattractant gradient as the population senses, engulfs and suppresses nutrient diffusion (darker regions indicate stronger chemoattractant concentration).

The low-concentration attractants used for the experiment in Fig. 3 presents a directed stimulus towards the collective (initialised at the bottom nutrient source). When the growing collective reaches the nearest source, its projection of attractant is damped by engulfment of the collective, and the collective splits to connect the left and right nutrients (whose distance from the current node is very similar). The suppression of each node as it is encountered - combined with the dynamically changing gradient field - presents a new stimulus pattern to the collective, until the spanning tree connecting all nutrients is formed.

We have also demonstrated using the model that it is possible to respond to differences in nutrient distance, nutrient size, and nutrient concentration by moving towards the closest, largest and highest concentration, respectively. The model also responds to the addition of new nutrients and the removal of nutrients (subject to a short delay whilst the diffusion stimulus profile is reconfigured) [5].

**Combining Attractant and Repellent Stimuli**

Attractant stimuli generate networks with proximity graph connectivity, for example Minimum Spanning Trees as in Fig. 3. In contrast, repellent stimuli generate plane division graphs as in the Voronoi diagram, where the model occupies the regions furthest away from the repellent sources (Fig. 4, left). By using individual repellent sources of varying concentrations the weighted Voronoi diagram can be approximated (Fig. 4, right). The combination of attractants and repellents allows us to guide the model using combinations of attractant and repellent stimuli (to 'pull' the collective towards certain regions or to 'push' the collective away from other regions). When similar concentrations of attractant and repellent stimuli are used there is no ambiguous behaviour. Instead hybrid graphs are formed (Fig. 5) which contain features of both proximity graphs and plane division graphs [6].

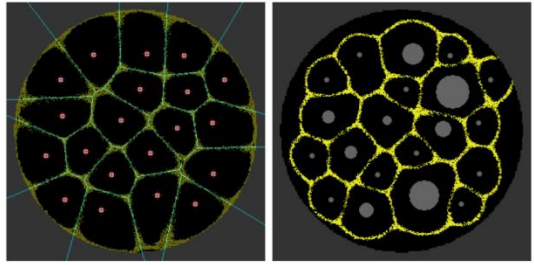

**Fig. 4:** Variation in repellent stimulus size allows approximation of the Weighted Voronoi diagram. Left: Repellent stimuli cause the model plasmodium to move away from stimuli sites forming Voronoi diagram (classically computed Voronoi diagram is overlaid). Right: Changing the size of the repellent stimuli causes greater distortions in the diffusion field, causing the model plasmodium to adopt patterns which approximate the multiplicatively weighted Voronoi diagram.

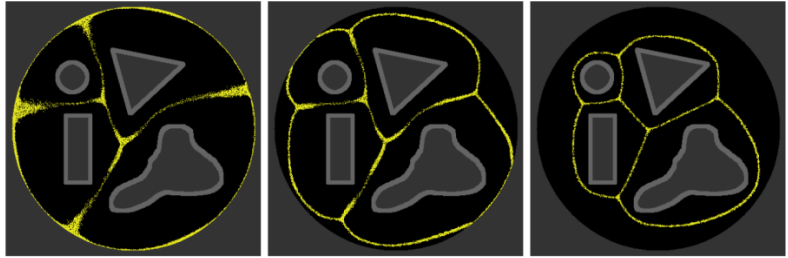

**Fig. 5:** Variation in repellent concentration cause a transition between classical Voronoi diagram and hybrid graphs which exhibit both plane division and minimisation features. Left: virtual plasmodium exposed to high-concentration repellent diffusion (from edge of planar shapes) adopts the Voronoi diagram. Middle: reducing the repellent concentration allows the contractile behaviour of the model to exert its effect in competition with the repellent field. Right: further reduction in repellent concentration forms a graph structure which exhibits both plane division and internal path minimisation features.

**Summary**

We demonstrate how the morphology of model slime mould transport networks is affected by a two-way interaction between the model plasmodium itself and the spatial stimuli generated by the parallel diffusion of attractants and repellents. It must be emphasised that the real-world stimuli presented to the *Physarum* plasmodium is much more complex (and noisy) than that presented to the model. However, the two-way mechanism is an efficient utilisation of the environment to 'outsource' part of the foraging computation used by the model. Variations in the concentration of attractants and repellents yield qualitatively different graph structures. Parallel propagation of information within a 2D environment is also exploited in chemical unconventional computing systems, however in these cases there is little scope for further modification of the diffusion stimuli. By sensing *and* dynamically remodelling the pattern of stimuli in its environment the slime mould performs a continuous adaptation to its environment. These results suggest novel forms of spatially represented unconventional computation using simple arrangements of external stimuli, coupled in a similar two-way interaction to morphologically adaptive collectives.

This work was supported by the EU research project "Physarum Chip: Growing Computers from Slime Mould" (FP7 ICT Ref 316366)

**References**

1. A. Adamatzky. *Physarum* Machines: Computers from Slime Mould, volume 74.World Scientific Pub Co Inc, 2010.
2. T. Nakagaki,, R. Kobayashi, Y. Nishiura & T. Ueda. Obtaining multiple separate food sources: behavioural intelligence in the *Physarum* plasmodium. Proceedings of the Royal Society of London. Series B: Biological Sciences, 271 (1554), 2305-2310. 2004
3. J. Jones. Characteristics of pattern formation and evolution in approximations of *Physarum* transport networks. Artificial Life, 16(2):127–153, 2010.
4. A. Takamatsu, E. Takaba & G. Takizawa. Environment-dependent morphology in plasmodium of true slime mold *Physarum polycephalum* and a network growth model. Journal of theoretical biology, 256(1), 29-44. 2009
5. J. Jones. Mechanisms Inducing Parallel Computation in a Model of *Physarum polycephalum* Transport Networks, Submitted, 2014.
6. J. Jones and A. Adamatzky. Slime Mould Inspired Generalised Voronoi Diagrams with Repulsive Fields, Int. Journal of Bifurcation and Chaos, In-press.